\documentclass[11pt]{article}
\usepackage[left=1in,top=1in,right=1in,bottom=1in]{geometry}
\usepackage{times}

\usepackage{verbatim}
\usepackage{amsmath}
\usepackage{amssymb}
\usepackage{amsthm}
\usepackage{rotating}
\usepackage{algorithm}
\usepackage[noend]{algpseudocode}

\usepackage{silence}
\usepackage{tabularx}
\usepackage{graphicx}
\usepackage{url}
\usepackage{multirow}
\usepackage{subfig}

\usepackage[noend]{algpseudocode}
\usepackage{siunitx}

\allowdisplaybreaks

\begin{document}
\title{Opponent Modeling in Multiplayer Imperfect-Information Games}
\author{Sam Ganzfried$^{1}$, Kevin A. Wang, Max Chiswick\\ 
$^1$Ganzfried Research, sam.ganzfried@gmail.com\\
}

\date{\vspace{-5ex}}

\maketitle

\begin{abstract}
In many real-world settings agents engage in strategic interactions with multiple opposing agents who can employ a wide variety of strategies. The standard approach for designing agents for such settings is to compute or approximate a relevant game-theoretic solution concept such as Nash equilibrium and then follow the prescribed strategy. However, such a strategy ignores any observations of opponents' play, which may indicate shortcomings that can be exploited. We present an approach for opponent modeling in multiplayer imperfect-information games where we collect observations of opponents' play through repeated interactions. We run experiments against a wide variety of real opponents and exact Nash equilibrium strategies in three-player Kuhn poker and show that our algorithm significantly outperforms all of the agents, including the exact Nash equilibrium strategies.  
\end{abstract}

\section{Introduction}
\label{se:intro}
Developing agents for strategic multiagent environments is challenging for several reasons. The classic game-theoretic setting assumes that there is no information available about the strategies of opposing agents. In this setting, the standard assumption is that all agents are behaving rationally, leading to the emergence of Nash equilibrium as the standard solution concept. While computing a Nash equilibrium can be done in polynomial time in two-player zero-sum games, it is PPAD-hard for non-zero-sum and multiplayer games and widely conjectured that no efficient algorithms exist~\cite{Chen05:Nash,Chen06:Settling,Daskalakis09:Complexity}. (Note that these results hold for extensive-form games of imperfect information as well as simultaneous strategic-form games.) In practice, we may have some information available on the opponents' strategies that we would like to take into account when determining our strategy. This information may include historical data of past play as well as observations from previous rounds within a repeated series of interactions. Within this framework there are many different settings which each may require different approaches. The historical data may be specific to our actual opponents, or may be from a general population from which our opponents are drawn. In the historical data, the private information of the agents may be revealed, not revealed, or partially revealed (i.e., revealed in certain observations and not in others). For real-time observations during repeated gameplay against the same opponents, the opponents' private information may also be fully, never, or partially revealed. 

In the general setting where historical data as well as real-time observations are available on our opponents, it seems natural to follow a Bayesian approach where we formulate a prior distribution over opponents' strategies based on the historical data which we update incrementally in real time as we obtain new observations. The general approach for accomplishing this in the two-player imperfect-information setting is depicted in Algorithm~\ref{al:meta}~\cite{Ganzfried18b:Bayesian}. This approach can be extended to multiple players by separately tracking the posterior distribution for each opposing player, then responding to the strategy profile consisting of the vector of the means of the individual posterior distributions. The key challenges of this approach are determination of appropriate prior strategy distributions and computation of (approximations of) the posterior distributions of the opponents' strategies. Even if we are able to satisfactorily accomplish these challenges, our algorithm may still fail in real-world environments because it implicitly assumes that the opponents are following static strategies. In general the opponents' strategies can be arbitrary and change over time, in which case the approach of building a posterior distribution based on observations of play over all timesteps may not provide a useful prediction of their strategies at the next time step. 

\begin{algorithm}[!ht]
\caption{Meta-algorithm for Bayesian opponent exploitation in two-player imperfect-information games~\cite{Ganzfried18b:Bayesian}}
\label{al:meta} 
\textbf{Inputs}: Prior distribution $p_0$, response functions $r_t$ for $0 \leq t \leq T$
\begin{algorithmic}
\State $M_0 \gets \overline{p_0(\sigma_{-i})}$
\State $R_0 \gets r_0(M_0)$
\State Play according to $R_0$
\For {$t = 1$ to $T$}
\State $x_t \gets $ observations of opponent's play at time step $t$
\State $p_t \gets$ posterior distribution of opponent's strategies given prior $p_{t-1}$ and observations $x_t$ 
\State $M_t \gets$ mean of $p_t$
\State $R_t \gets r_t(M_t)$
\State Play according to $R_t$
\EndFor
\end{algorithmic}
\end{algorithm}

In this work we study the setting in which no historical data is available and there is partial observability of opponents' private information, which is perhaps the most challenging setting (note that an approach for the partial observability setting must also be able to handle the extreme cases of full or no observability). We experiment in a three-player imperfect-information game against a set of real opponent agents who can be playing arbitrary strategies that may be dynamic. We also include several exact Nash equilibrium strategies as opponents in our experiments. To obtain a strong performance overall, our algorithm must exploit the weaker opposing agents while still performing relatively well against the stronger opponents. In addition to the assumption that no historical data is available, our approach also does not assume access to any domain-specific knowledge or features. So our approach is fully domain-independent and applicable to any multiplayer imperfect-information game in the absence of expertise of the domain. Our approach is also scalable to large games. We demonstrate that we are able to obtain a significantly higher payoff than the Nash equilibrium agents in our setting against a wide variety of real agents. 

\section{Comparison to related research}
\label{se:related}
There have been several prior approaches for opponent modeling, though typically these are for different settings, and in particular most of the approaches are just for two-player zero-sum games. An early work applied a Bayesian approach to two-player zero-sum imperfect-information games (and variants of two-player poker specifically)~\cite{Southey05:Bayes}. That approach also addresses the setting in which no historical data is available and there is partial observability of the opponent's private information (though there is just a single opponent). They consider three different approaches for responding to the posterior model of the opponent's strategies called Bayesian Best Response, Max A Posteriori Response, and Thompson's Response. All of these approaches involve sampling and are scalable to large games (though provide only approximations of the posterior of the opponent's strategy). Subsequent research on the Bayesian setting has shown that the Bayesian Best Response significantly outperforms the other two approaches when compared against an exact (unsampled) posterior best response in a simplified setting~\cite{Ganzfried18b:Bayesian}. So our approach will be based on the Bayesian Best Response approach. The original approach used two different prior models for the opponent's strategy. The first is a model that assumes independent Dirichlet distributions for each information set, which are na\"ively set to have all parameters equal to 2 and therefore have the strategy that selects each action with equal probability as its mean. The second is an ``informed prior'' that assumes five different game-specific features determined by consultation with a domain expert. While our general approach is similar, there are several key differences. First, this approach is just for two-player zero-sum games, and our approach creates opponent models for any number of opposing agents. And second, our approach considers an independent Dirichlet prior model whose mean is a Nash equilibrium strategy, not a na\"ive random strategy. While we are considering a game that is small enough that exact Nash equilibrium computation is tractable, there is still a further challenge of deciding which Nash equilibrium to use as the prior as our game contains infinitely many. For larger multiplayer games in which exact Nash equilibrium computation is intractable we can use an approximation of one instead. 

A more recent Bayesian approach has been developed for the two-player imperfect-information zero-sum setting with no observability of the opponent's private information~\cite{Ganzfried18b:Bayesian}. An exact (unsampled) Bayesian best response approach was devised for this setting, which was demonstrated to outperform the three approximation approaches described above from prior work~\cite{Southey05:Bayes}. However, this approach does not deal with the setting of partial observability, also assumes a na\"ive independent Dirichlet prior distribution with all parameter values equal, and furthermore is not scalable to large games. 

In contrast to the approaches that use a na\"ive independent Dirichlet prior, the deviation-based best response approach uses an approximation of Nash equilibrium strategies as the prior mean~\cite{Ganzfried11:Game}. The approach incorporates behavioral strategy constraints and finds a model for the opponent's strategy that is closest to the precomputed Nash equilibrium approximation subject to being consistent with our observations of their play. This approach is for the two-player imperfect-information zero-sum setting with no observability of the opponent's private information and is not directly applicable to the setting of partial observability that we are considering. In many real-world settings, including poker, the opponents' private information is only revealed in certain situations; for example, in poker we only see the opponents' private cards if the hand goes to a ``showdown,'' and we do not see their cards if they fold. This approach was demonstrated to outperform an approximate Nash equilibrium strategy against several of the worst-performing agents in limit Texas hold 'em from the Annual Computer Poker Competition. However, the approach was not tested against strong or medium-strength opponents. In our experiments we will experiment against opponents of a wide range of abilities, including several exact Nash equilibrium strategies. While outperforming Nash equilibrium against very weak opponents is a good start, it is far from the end goal for opponent modeling. In the real world we want to obtain strong performance against a wide range of potential opponents, some of whom may be quite strong and potentially deceptive. 

The paradigm of incorporating behavioral strategy constraints was also utilized to create an agent for opponent modeling in the small game of two-player Leduc hold 'em~\cite{Davis19:Solving}. This approach first collects data on the opponent's strategy by following a ``probe'' strategy that follows a uniform random distribution over non-fold options. This probe strategy is utilized for a large number of iterations to generate observations of the opponent's play from which we can construct strategy constraints. Then an algorithm is run on the constrained game to construct a counterstrategy to the opponent. While this approach is scalable and applies to the setting of partial observability, it assumes access to a significant amount of prior information on the opponent's strategy that is learned during the probing period. If we were to follow a random probing strategy for a large number of iterations in practice, the poor performance during the probing period would almost certainly offset any performance improvement of opponent modeling unless we play against the same opponent for an extremely large number of iterations.

Bard et al. take a different approach that precomputes several exploitative strategies offline and selects between them online using the Exp4 multi-armed bandit algorithm~\cite{Bard13:Online}. To construct the precomputed strategies, they have access to a database of over 8.4 million hands with full observability of the agents' private information. From this database they construct approximate response strategies to several of these agent strategies in an abstracted version of the game, and construct a portfolio from these response strategies. The approach achieved strong performance against agents in limit Texas hold 'em (where the experiments were against different agents than those in the historical database).  Note that we are interested in the setting where no historical data is available, and so this approach is not applicable. 

To summarize, we study the setting where there can be any number of opponents, there is no historical data available (on any agents, not just on our opponents), there is partial observability of opponents' private information (which includes as special cases full and no observability), and we have no access to domain-specific information such as features. Our goal is to obtain a strong performance against a wide range of unknown opponents within a limited number of repeated interactions. We cannot simply follow a random ``probe'' strategy in order to learn our opponents' strategies, since the poor performance obtained during the probing period will offset the benefits of improved opponent modeling. Instead we start by following a Nash equilibrium strategy until we have collected enough observations to start effectively exploiting the opponents. While this may not collect observations as efficiently as a random probing strategy, it will obtain a strong performance throughout the initial exploration period. Our approach uses importance sampling to estimate the posterior mean of the opponents' strategies and is scalable to large games. Our approach is domain independent and assumes only access to partial observations of opponents' play in the current series of games. We note that if additional information is available such as a historical database of play or domain-specific features, naturally we would want to incorporate this information into our approach to obtain stronger performance. However often such information is not available, and we seek to obtain a strong approach that does not rely on them. If such information is available, our approach can still be used as a strong starting point which can be augmented by integrating this additional information.

\section{Algorithm}
\label{se:algorithm}
A \emph{strategic-form game} (aka \emph{normal-form game}) consists of a finite set of players $N = \{1,\ldots,n\}$, a finite set of pure strategies $S_i$ for each player $i \in N$, and a real-valued utility for each player for each strategy vector (aka \emph{strategy profile}), $u_i : \times_i S_i \rightarrow \mathbb{R}$. A \emph{mixed strategy} $\sigma_i$ for player $i$ is a probability distribution over pure strategies, where $\sigma_i(s_{i'})$ is the probability that player $i$ plays pure strategy $s_{i'} \in S_i$ under $\sigma_i$. Let $\Sigma_i$ denote the full set of mixed strategies for player $i$. A strategy profile $\sigma^* = (\sigma^*_1,\ldots,\sigma^*_n)$ is a \emph{Nash equilibrium} if $u_i(\sigma^*_i,\sigma^*_{-i}) \geq u_i(\sigma_i, \sigma^*_{-i})$ for all $\sigma_i \in \Sigma_i$ for all $i \in N$, where $\sigma^*_{-i} \in \Sigma_{-i}$ denotes the vector of the components of strategy $\sigma^*$ for all players excluding player $i$. Here $u_i$ denotes the expected utility for player $i$, and $\Sigma_{-i}$ denotes the set of strategy profiles for all players excluding player $i$. While the strategic form can be used to model simultaneous actions, settings with sequential moves are typically modelled using the \emph{extensive form} representation. 
Extensive-form games consist primarily of a game tree; each non-terminal node has an associated player (possibly \emph{chance}) that makes the decision at that node, and each terminal node has associated utilities for the players.  Additionally, game states are partitioned into \emph{information sets}, where the player whose turn it is to move cannot distinguish among the states in the same information set.  Therefore, in any given information set, a player must choose actions with the same distribution at each state contained in the information set. If no player forgets information that they previously knew, we say that the game has \emph{perfect recall}. A \emph{strategy} for player $i,$ $\sigma_i \in \Sigma_i,$ is a function that assigns a probability distribution over all actions at each information set belonging to $i$. Similarly to strategic-form games, it can be shown that all extensive-form games with perfect recall contain at least one Nash equilibrium. In this section we present our algorithm for opponent modeling in multiplayer extensive-form games of imperfect information.

We study the setting where the original \emph{stage game} is repeated for $T$ iterations. After the completion of each iteration, private information of some players may be observed by other players depending on the information revelation structure of the game. Between game iterations the positions of the players in the game may change. For example, in three-player poker, the player first to act in iteration $t$ is third to act in iteration $t+1$, the second player to act in iteration $t$ is first to act in iteration $t+1$, and the third player to act in iteration $t$ is the second to act in iteration $t+1.$ Thus we must differentiate the player label from the player position in the original game, where the player position corresponds to the ``player'' in the original game definition. For simplicity assume that our player label is 1. In the three-player poker setting we must create six different opponent models: for player 2 when we are in positions 1--3, and for player 3 when we are in positions 1--3. (Note that after each iteration of play only two of these models are updated.) We assume that all six of these strategies are independent, and furthermore that each strategy selects its action independently at each information set.

Our general opponent modeling approach is described in Algorithm~\ref{al:opp-model}. At a high level, the algorithm works as follows. Given a threshold $H < T,$ for the first $H$ iterations we follow our ``default'' strategy $\sigma^*$, while collecting observations of opponents' play. In the absence of any additional information, we will choose $\sigma^*$ to be an exact or approximate Nash equilibrium strategy. We also assume that we have access to default strategies $\{\sigma_{i,j}\}$ for all opposing player label/position combinations to use as the means of the prior distribution for the strategy. In the absence of additional information, we will again choose these to be exact or approximate Nash equilibria. Note that games may have multiple Nash equilibrium strategies and we may prefer some over others (and furthermore we may prefer to use one for our default strategy that differs from the ones used for the opponent priors). Prior to game play, we simulate $k$ strategies for each opponent player label/position combination by sampling from a distribution with mean at $\sigma_{i,j}.$ We will be using Dirichlet distributions for the priors, which is the most commonly-used prior distribution in this context. The prior distributions will use a multiplicative scaling factor parameter $\eta$ as well as a rounding parameter $\epsilon$ to ensure that no actions have probability 0 in the prior. We initialize all priors over joint sampled opponent strategies to have equal probability. After each iteration, we update these probabilities by applying Bayes' rule given our observations to compute the posterior probabilities of the sampled strategies. After we reach iteration $H$, we compute and play a best response to the joint model of the opponents' strategies at each iteration until we reach the final iteration $T.$

\begin{algorithm}
\caption{Opponent modeling algorithm for multiplayer imperfect-information games}
\label{al:opp-model}
\textbf{Inputs}: Rounding threshold $\epsilon$, number of samples $k$, switching time $H$, our default strategy $\sigma^*$, prior strategy means $\{\sigma_{i,j}\}$ for all opposing players $i$ in position $j$, Dirichlet factor parameter $\eta.$
\begin{algorithmic}
\State Simulate $k$ strategies for each opponent $i$ in each position $j$ by $\{\tau_{i,j,s}\}$ = CreateSamples($\epsilon, k, \{\sigma_{i,j}\}, \eta$).
\State Initialize prior probabilities of all samples to be equal.
\For {$t = 1$ to $H$}
\State Follow $\sigma^*$ and collect observations on opponents' play.
\State Update posterior strategy probabilities from the new observations to create opponent models.
\EndFor
\For {$t=H+1$ to $T$}
\State Update posterior strategy probabilities from the new observations to create opponent models.
\State Compute and play best response strategy to the opponent models.
\EndFor
\end{algorithmic}
\end{algorithm}

The procedure for creating the sampled strategies for each opponent player label/position is given in Algorithm~\ref{al:create-samples}. For each player label $i$ and position $j$, we generate $k$ samples from a Dirichlet distribution with mean $\sigma_{i,j}.$ Since we are using an independent Dirichlet prior model, we assume that the strategies at each information set follow independent Dirichlet distributions. Before constructing the samples we ensure that there are no probability zero actions in the prior mean by setting action probabilities below $\epsilon$ to $\epsilon$ and probabilities above $1-\epsilon$ to $1-\epsilon$ then normalizing. The term $\sigma^{q_{i,j}}_{i,j}(a)$ denotes the probability that mixed strategy $\sigma_{i,j}$ selects action $a$ at information set $q_{i,j}.$ After performing this rounding and renormalization, we then generate $k$ samples independently at each information set from Dirichlet distributions. Note that for a Dirichlet distribution $f_{Dir}(x_1,\ldots,x_K; \alpha_1,\ldots,\alpha_K)$, we have $E[X_i] = \frac{\alpha_i}{\sum_{i=1}^K \alpha_i}.$ Therefore, the mean remains the same if we multiply all parameters by a constant. This constant, which we denote by $\eta,$ is a parameter of the algorithm. It is well-known that to generate a sample from a Dirichlet distribution with parameters $(\alpha_1,\ldots,\alpha_K)$ we first draw $K$ independent samples $y_1,\ldots,y_K$ from Gamma distributions each with density Gamma$(\alpha_i,1) = \frac{y^{\alpha_i-1}_i - e^{-y_i}}{\Gamma(\alpha_i)}$ and then set $x_i = \frac{y_i}{\sum_{j=1}^K y_j}$~\cite{Wiki22:Dirichlet}.

\begin{algorithm}
\caption{CreateSamples}
\label{al:create-samples}
\textbf{Inputs}: Rounding threshold $\epsilon$, number of samples $k$, prior strategy means $\{\sigma_{i,j}\}$ for all opposing players $i$ in position $j$, Dirichlet factor parameter $\eta.$
\begin{algorithmic}
\For {each opposing player $i$ and position $j$}
\For {each information set $q_{i,j}$}
\For {each action $a$ at information set $q_{i,j}$}
\If {$\sigma^{q_{i,j}}_{i,j}(a) < \epsilon$}
\State $\sigma^{q_{i,j}}_{i,j}(a) = \epsilon$
\ElsIf {$\sigma^{q_{i,j}}_{i,j}(a) > 1 -\epsilon$}
\State $\sigma^{q_{i,j}}_{i,j}(a) = 1 -\epsilon$
\EndIf
\EndFor
\State Normalize $\sigma^{q_{i,j}}_{i,j}$
\For {$s = 1$ to $k$}
\State $z_s = 0$
\For {each action $a$ at information set $q_{i,j}$}
\State $x_a$ = sample from Gamma distribution with parameters $\eta \cdot \sigma^{q_{i,j}}_{i,j}(a)$ and 1
\State $z_s = z_s + x_a$
\EndFor
\For {each action $a$ at information set $q_{i,j}$}
\State $\tau^{q_{i,j}}_{i,j,s}(a) = \frac{x_a}{z_s}$
\EndFor
\EndFor
\EndFor
\EndFor
\Return $\{\tau_{i,j,s}, s = 1 \mbox{ to } k\}$
\end{algorithmic}
\end{algorithm}

The procedure for updating the posterior strategy probabilities is given in Algorithm~\ref{al:update-posteriors}. Note that we only need to update these probabilities for the strategy components for the specific player positions the opponents were in during the previous iteration, of which there are $k$ for each player. For each combination of these samples, we apply Bayes' rule to compute the posterior probability proportional to the product of the previous posterior probability of the sample combination (which serves as the current prior) and the probability that we would have observed play consistent with our observations $o$ if we followed our current strategy $\sigma$ and the opponents follow their corresponding sampled strategies. Note that the observations $o$ capture partial observability of opponents' private information as determined by the information revelation structure of the game. For example, in poker all actions are publicly observable and the private cards of an opponent are visible only during hands where the opponent went to ``showdown.'' (Of course, our own private cards are always observable by ourselves.) A showdown in poker occurs when at least two players have not folded before the end of the betting, in which case the players must ``show'' their cards to reveal who has the highest hand. (By contrast, if all players except one fold during the hand, the player who did not fold wins the pot without having to reveal their hand.)

\begin{algorithm}
\caption{Update posterior strategy probabilities}
\label{al:update-posteriors}
\textbf{Inputs}: Current posterior probabilities $p(s_1,\ldots,s_n)$ for sample $s_i$ for opposing player $i$, sampled strategies $\{\tau_{i,j,s}\}$ for all opposing players, our current strategy $\sigma$, observations from current round of play $o.$
\begin{algorithmic}
\State $z = 0$
\For {each combination of sample indices $s = (s_1,\ldots,s_n)$}
\State $z_s$ = probability we observe play consistent with $o$ when we follow $\sigma$ and opponents follow sampled strategies $\tau_{1,j_1,s_1},\ldots,\tau_{n,j_n,s_n}$
\State $q(s_1,\ldots,s_n) = p(s_1,\ldots,s_n) * z_s$
\State $z = z + p(s_1,\ldots,s_n) * z_s$
\EndFor
\For {each combination of sample indices $s = (s_1,\ldots,s_n)$}
\State $q(s_1,\ldots,s_n) = \frac{q(s_1,\ldots,s_n)}{z}$
\EndFor
\Return {$q(s_1,\ldots,s_n)$}
\end{algorithmic}
\end{algorithm}

The final subroutine of computing the opponent models from the newly-computed posterior probabilities is given in Algorithm~\ref{al:opponent-models}. This follows the Bayesian Best Response approach of Southey et al.~\cite{Southey05:Bayes}, which has been demonstrated to significantly outperform the Max A Posteriori Response and Thompson's Response approaches for Bayesian opponent modeling~\cite{Ganzfried18b:Bayesian}. As for updating the posteriors, we only need to compute new opponent models for the strategy components for the specific player positions the opponents were in during the previous iteration. The term $m^{q_{i,j}}_{i,j}(a)$ denotes the new model for the probability that opponent player label $i$ in position $i_j$ takes action $a$ at information set $q_{i,j}$. This is set to be the convex combination of the sampled strategies where the weights are the newly-computed posterior probabilities.

\begin{algorithm}
\caption{Compute opponent models}
\label{al:opponent-models}
\textbf{Inputs}: Current posterior probabilities $p(s_1,\ldots,s_n)$ for sample $s_i$ for opposing player with label $i$ and position $i_j$, sampled strategies $\{\tau_{i,j,s}\}$ for all opposing player labels $i$ and positions $i_j$
\begin{algorithmic}
\State $m_{i,j} \gets$ vector of all zeros for all info. sets and actions.
\For {opposing players $i$ in position $i_j$}
\For {each information set $q_{i,j}$}
\For {each action $a$ at information set $q_{i,j}$}
\For {each combination of sample indices $s = (s_1,\ldots,s_n)$}
\State $m^{q_{i,j}}_{i,j}(a) = m^{q_{i,j}}_{i,j}(a) + p(s_1,\ldots,s_n)\tau^{q_{i,j}}_{i,j,s_i}(a)$
\EndFor
\EndFor
\EndFor
\EndFor
\Return {$\{m_{i,j}\}$}
\end{algorithmic}
\end{algorithm}

Overall our algorithm consists of four parameters---$\epsilon, k, H, \eta$---as well as our default strategy for all player positions $\sigma^*$ and prior strategy means for all opposing player label/position combinations $\{\sigma_{i,j}\}$, which can be viewed as additional parameters. The algorithm applies to all multiplayer imperfect information games with partial observability of private information. The running time of the algorithm is $O((n-1)k^{n-1}|I|^n|A|^n)$, where $k$ is the number of sampled strategies for each player label/position combination, $n$ is the number of players, $|I|$ is the max number of information sets per player, and $|A|$ is the max number of actions per player. Since game representations are $O(n |I|^n |A|^n)$, this algorithm is efficient with respect to the size of the game representation (assuming $n$ is a constant) and is scalable to large games.

\section{Three-player Kuhn poker}
\label{se:kuhn}
We will experiment with our algorithm against a variety of agents in three-player Kuhn poker. 
Three-player Kuhn poker is a simplified form of limit poker that has been used as a testbed game in the AAAI Annual Computer Poker Competition for several years. There is a single round of betting. Each player first antes a single chip and is dealt a card from a four-card deck that contains one Jack (J), one Queen (Q), one King (K), and one Ace (A). The first player has the option to \emph{bet} a fixed amount of one additional chip (by contrast in \emph{no-limit} games players can bet arbitrary amounts of chips) or to \emph{check} (remain in the hand but not bet an additional chip). When facing a bet, a player can \emph{call} (i.e., match the bet) or \emph{fold} (forfeit the hand). No additional bets or raises beyond the additional bet are allowed (while they are allowed in other common poker variants such as Texas hold 'em). If all players but one have folded, then the player who has not folded wins the \emph{pot}, which consists of all chips in the middle. If more than one player has not folded by the end there is a \emph{showdown}, at which the players reveal their private card and the player with the highest card wins the entire pot (which consists of the initial antes plus all additional bets and calls). The Ace is the highest card, followed by King, Queen, and Jack. As one example of a play of the game, suppose the players are dealt Queen, King, Ace respectively, and player 1 checks, player 2 checks, player 3 bets, player 1 folds, and player 2 calls; then player 3 would win a pot of 5, for a profit of 3 from the start of the hand (while player 1 loses 1 and player 2 loses 2).

Note that despite the fact that 3-player Kuhn poker is only a synthetic simplified form of poker and is not actually played competitively, it is still far from trivial to analyze, and contains many of the interesting complexities of popular forms of poker such as Texas hold 'em. First, it is a game of imperfect information, as players are dealt a private card that the other agents do not have access to, which makes the game more complex than a game with perfect information that has the same number of nodes. Despite the size, it is not trivial to compute Nash equilibrium analytically, though recently an infinite family of Nash equilibria has been computed~\cite{Szafron13:Parametrized}. The equilibrium strategies exhibit the phenomena of \emph{bluffing} (i.e., sometimes betting with weak hands such as a Jack or Queen), and \emph{slow-playing} (aka \emph{trapping}) (i.e., sometimes checking with strong hands such as a King or Ace in order to induce a bet from a weaker hand). 

A full infinite family of Nash equilibria for this game has been computed and can be seen in the tables from a recent article by Szafron et al.~\cite{Szafron13:Parametrized}. The family of equilibria is based on several parameter values, which once selected determine the probabilities for the other portions of the strategies. One can see that randomization and including some probability on trapping and bluffing are essential in order to have a strong and unpredictable strategy. Thus, while this game may appear quite simple at first glance, analysis is still very far from simple, and the game exhibits many of the complexities of far larger games that are played competitively by humans for large amounts of money. 

\section{Experiments}
\label{se:experiments}
We experimented with our agent against ten agents created by students for a class project as well as three exact Nash equilibrium strategies. The class agents exhibit a wide range of abilities and utilize a wide range of approaches that are potentially dynamic. (These approaches include neural networks, counterfactual regret minimization, opponent modeling, and rule-based approaches.) The Nash equilibrium strategies are all from the ``robust'' subfamily of equilibrium strategies that have been singled out as obtaining the best worst-case payoff assuming that the other agents are following one of the strategies given by the computed infinite equilibrium family~\cite{Szafron13:Parametrized}. In particular, we use the strategy falling at the lower bound of this robust equilibrium subfamily, the upper bound, and the midpoint. For each grouping of 3 agents we ran 10 matches consisting of 3000 hands between each of the 6 permutations of the agents (with the same cards being dealt for the respective positions of the agents in each of the duplicated matches). The number of hands per match (3000) is the same value used in the Annual Computer Poker Competition, and the process of duplicating the matches with the same cards between the different agent permutations is a common approach that significantly reduces the variance.

For our agent we use the following parameter values: $\epsilon = 0.05$, $k = 10$, $H = 100$, $\eta = 4.$ For our default strategy $\sigma^*$ we use the strategy at the lower bound of the space of robust Nash equilibrium strategies, and for the prior strategy means for the opponents $\{\sigma_{i,j}\}$ we use the strategy at the midpoint of the space of robust Nash equilibrium strategies. We chose this strategy for $\sigma^*$ because it performed the best out of the lower bound, upper bound, and midpoint in preliminary experiments. We chose the midpoint as the prior means for the opponents because it has the most entropy out of the three strategies and has fewer actions with probability 0 which may be problematic when modeling unknown opponents. The results from these experiments are shown in Table~\ref{ta:results1}. MBBR denotes our multiplayer Bayesian best response agent, N1 denotes the lower bound Nash agent, N2 denotes the upper bound Nash agent, N3 denotes the midpoint Nash agent, and C1--10 are the class project agents. The values are the winrates of the agents per hand multiplied by 1,000 (i.e., a winrate of 30 means winning 0.03 chips per hand, or 30 \emph{millichips} per hand). The standard errors for all agents are between 0.4 and 0.6 (using the same units). The MBBR agent clearly outperformed all other agents with statistical significance. We can see that the exact Nash equilibrium agents also performed well, coming in 2nd, 3rd, and 6th place. The strategies for the Nash agents N1, N2, and N3 are depicted in Tables~\ref{ta:ne1}--\ref{ta:ne3}. The tables assign values for the 21 free parameters in the infinite family of Nash equilibrium strategies~\cite{Szafron13:Parametrized}. To define these parameters, $a_{jk}$, $b_{jk}$, and $c_{jk}$ denote the action probabilities for players 1, 2, and 3 respectively when holding card $j$ and taking an aggressive action (Bet (B) or Call (C)) in  situation $k$, where the betting situations are defined in Table~\ref{ta:betting} (K denotes check and F denotes fold).

\begin{table}[!ht]
\centering
\begin{tabular}{|*{14}{c|}} \hline
MBBR &N1 &N3 &C1 &C2 &N2 &C3 &C4 &C5 &C6 &C7 &C8 &C9 &C10\\ \hline
48 &37 &35 &34 &33 &30 &22 &14 &11 &7 &-22 &-33 &-44 &-172\\ \hline
\end{tabular}
\caption{Experimental results against Nash and class agents.}
\label{ta:results1}
\end{table}
\normalsize
\renewcommand{\tabcolsep}{6pt}

\begin{table}[!ht]
\centering
\begin{tabular}{|*{3}{c|}} \hline
P1 & P2 & P3\\ \hline
$a_{11} = 0$ & $b_{11} = 0$ & $c_{11} = 0$ \\ \hline
$a_{21} = 0$ & $b_{21} = 0$ & $c_{21} = \frac{1}{2}$ \\ \hline
$a_{22} = 0$ & $b_{22} = 0$ & $c_{22} = 0$ \\ \hline
$a_{23} = 0$ & $b_{23} = 0$ & $c_{23} = 0$ \\ \hline
$a_{31} = 0$ & $b_{31} = 0$ & $c_{31} = 0$ \\ \hline
$a_{32} = 0$ & $b_{32} = 0$ & $c_{32} = 0$ \\ \hline
$a_{33} = \frac{1}{2}$ & $b_{33} = \frac{1}{2}$ & $c_{33} = \frac{1}{2}$ \\ \hline
$a_{34} = 0$ & $b_{34} = 0$ & $c_{34} = 0$ \\ \hline
$a_{41} = 0$ & $b_{41} = 0$ & $c_{41} = 1$ \\ \hline
\end{tabular}
\caption{Parameter values used for Nash agent N1.}
\label{ta:ne1}
\end{table}

\begin{table}[!ht]
\centering
\begin{tabular}{|*{3}{c|}} \hline
P1 & P2 & P3\\ \hline
$a_{11} = 0$ & $b_{11} = \frac{1}{4}$ & $c_{11} = 0$ \\ \hline
$a_{21} = 0$ & $b_{21} = \frac{1}{4}$ & $c_{21} = \frac{1}{2}$ \\ \hline
$a_{22} = 0$ & $b_{22} = 0$ & $c_{22} = 0$ \\ \hline
$a_{23} = 0$ & $b_{23} = 0$ & $c_{23} = 0$ \\ \hline
$a_{31} = 0$ & $b_{31} = 0$ & $c_{31} = 0$ \\ \hline
$a_{32} = 0$ & $b_{32} = 1$ & $c_{32} = 0$ \\ \hline
$a_{33} = \frac{1}{2}$ & $b_{33} = \frac{7}{8}$ & $c_{33} = 0$ \\ \hline
$a_{34} = 0$ & $b_{34} = 0$ & $c_{34} = 1$ \\ \hline
$a_{41} = 0$ & $b_{41} = 1$ & $c_{41} = 1$ \\ \hline
\end{tabular}
\caption{Parameter values used for Nash agent N2.}
\label{ta:ne2}
\end{table}

\begin{table}[!ht]
\centering
\begin{tabular}{|*{3}{c|}} \hline
P1 & P2 & P3\\ \hline
$a_{11} = 0$ & $b_{11} = \frac{1}{8}$ & $c_{11} = 0$ \\ \hline
$a_{21} = 0$ & $b_{21} = \frac{1}{8}$ & $c_{21} = \frac{1}{2}$ \\ \hline
$a_{22} = 0$ & $b_{22} = 0$ & $c_{22} = 0$ \\ \hline
$a_{23} = 0$ & $b_{23} = 0$ & $c_{23} = 0$ \\ \hline
$a_{31} = 0$ & $b_{31} = 0$ & $c_{31} = 0$ \\ \hline
$a_{32} = 0$ & $b_{32} = \frac{23}{64}$ & $c_{32} = 0$ \\ \hline
$a_{33} = \frac{1}{2}$ & $b_{33} = \frac{11}{16}$ & $c_{33} = \frac{1}{4}$ \\ \hline
$a_{34} = 0$ & $b_{34} = 0$ & $c_{34} = \frac{1}{2}$ \\ \hline
$a_{41} = 0$ & $b_{41} = \frac{1}{2}$ & $c_{41} = 1$ \\ \hline
\end{tabular}
\caption{Parameter values used for Nash agent N3.}
\label{ta:ne3}
\end{table}

\begin{table}[!ht]
\centering
\begin{tabular}{|*{4}{c|}} \hline
Situation & P1 & P2 & P3\\ \hline
1 & -- & K & KK \\ \hline
2 & KKB & B & KB \\ \hline
3 & KBF & KKBF & BF \\ \hline
4 & KBC & KKBC & BC \\ \hline
\end{tabular}
\caption{Betting situations in three-player Kuhn poker.}
\label{ta:betting}
\end{table}

\begin{figure}[!ht]
\centering
\includegraphics[scale = 0.58]{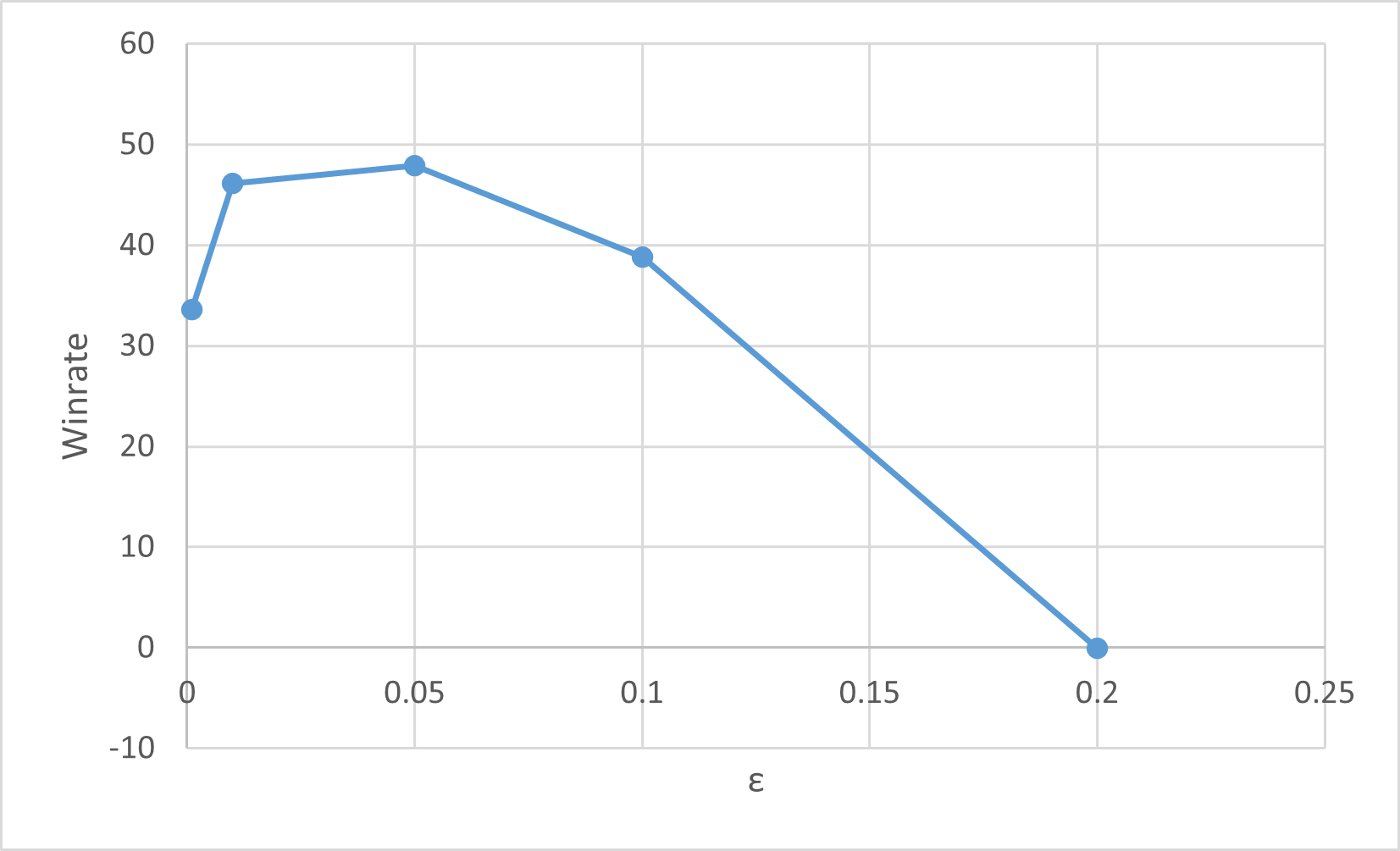}
\caption{Winrate in millichips per hand as a function of $\epsilon.$}
\label{fi:epsilon}
\end{figure}

\begin{figure}[!ht]
\centering
\includegraphics[scale = 0.58]{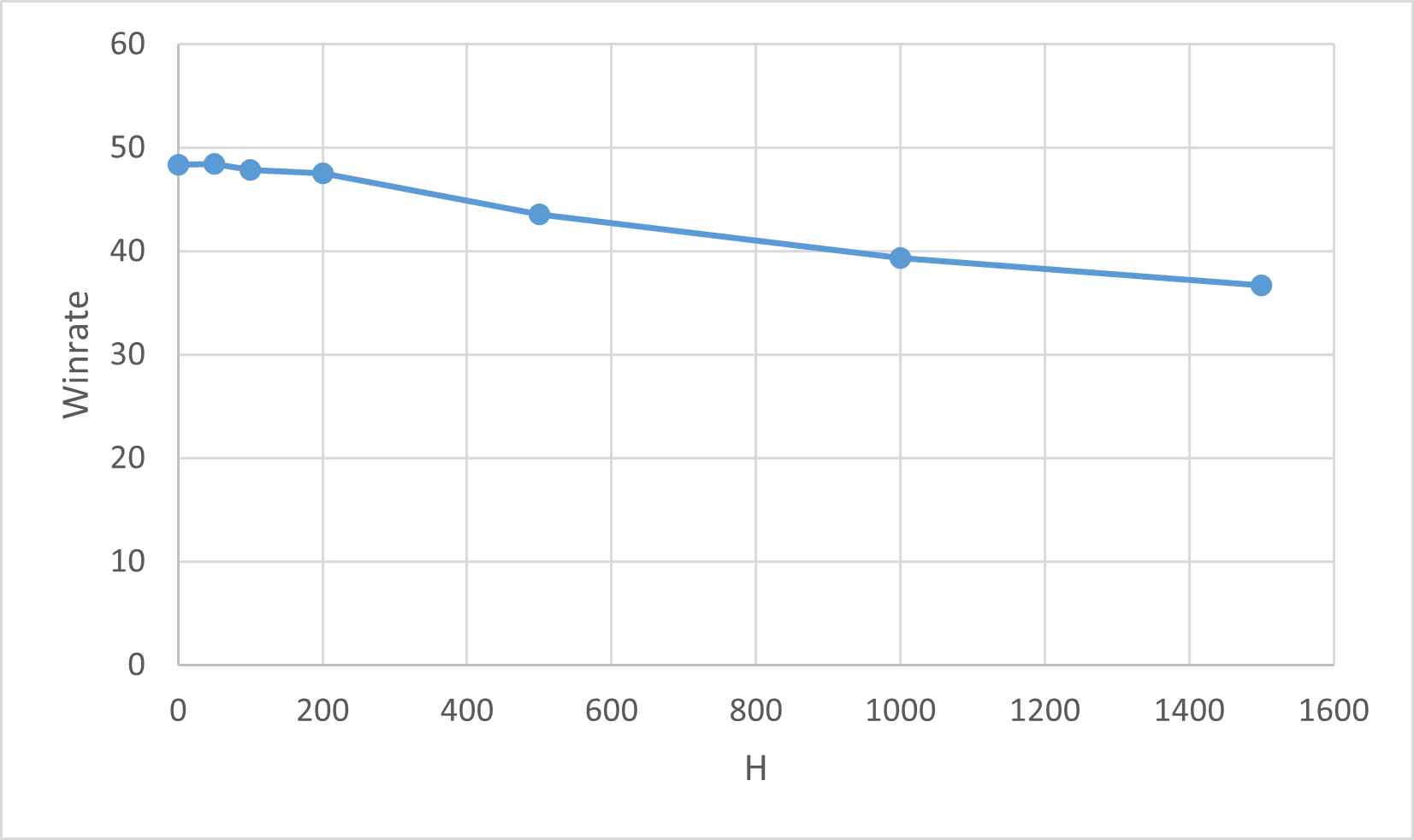}
\caption{Winrate in millichips per hand as a function of $H.$}
\label{fi:threshold}
\end{figure}

\begin{figure}[!ht]
\centering
\includegraphics[scale = 0.58]{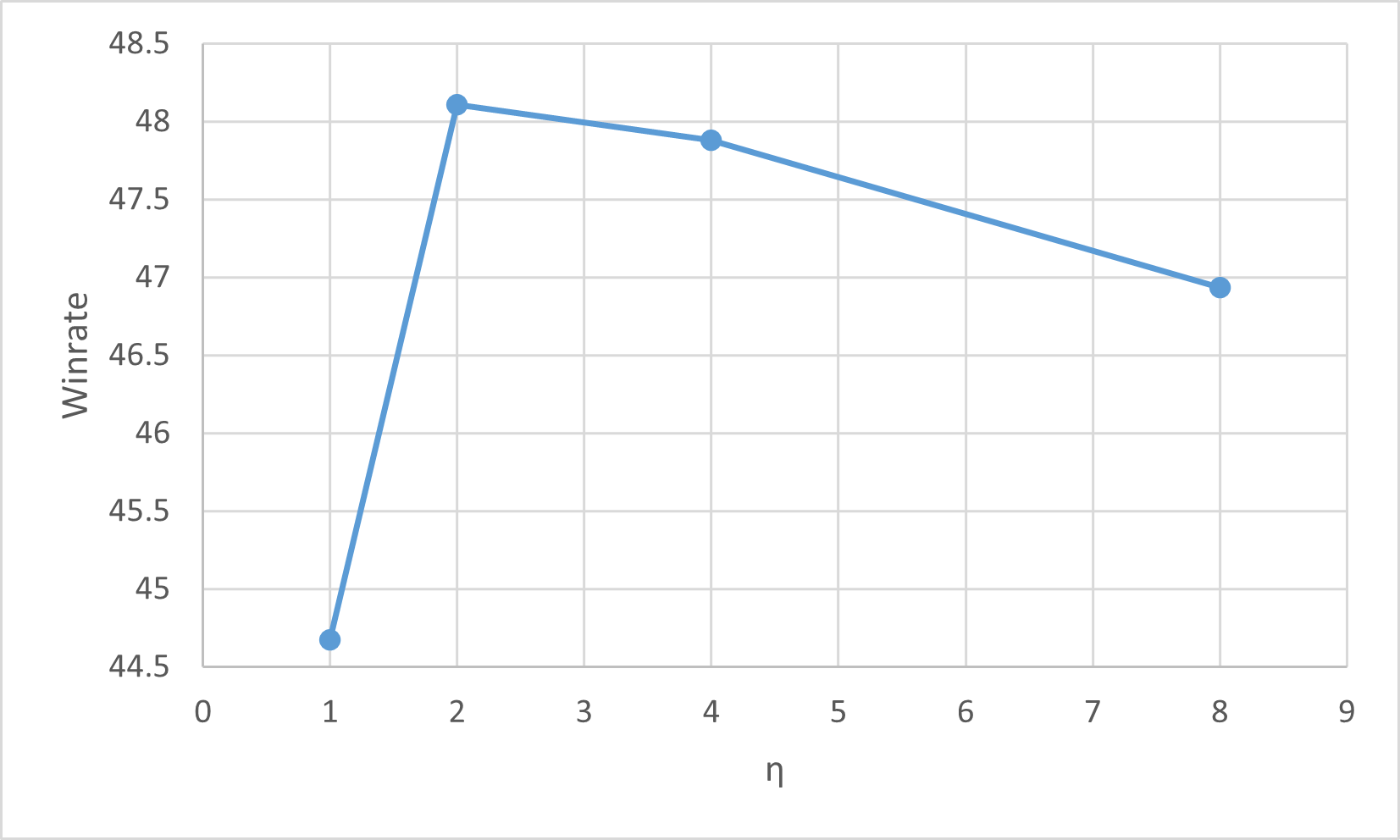}
\caption{Winrate in millichips per hand as a function of $\eta.$}
\label{fi:factor}
\end{figure}

\begin{figure}[!ht]
\centering
\includegraphics[scale = 0.58]{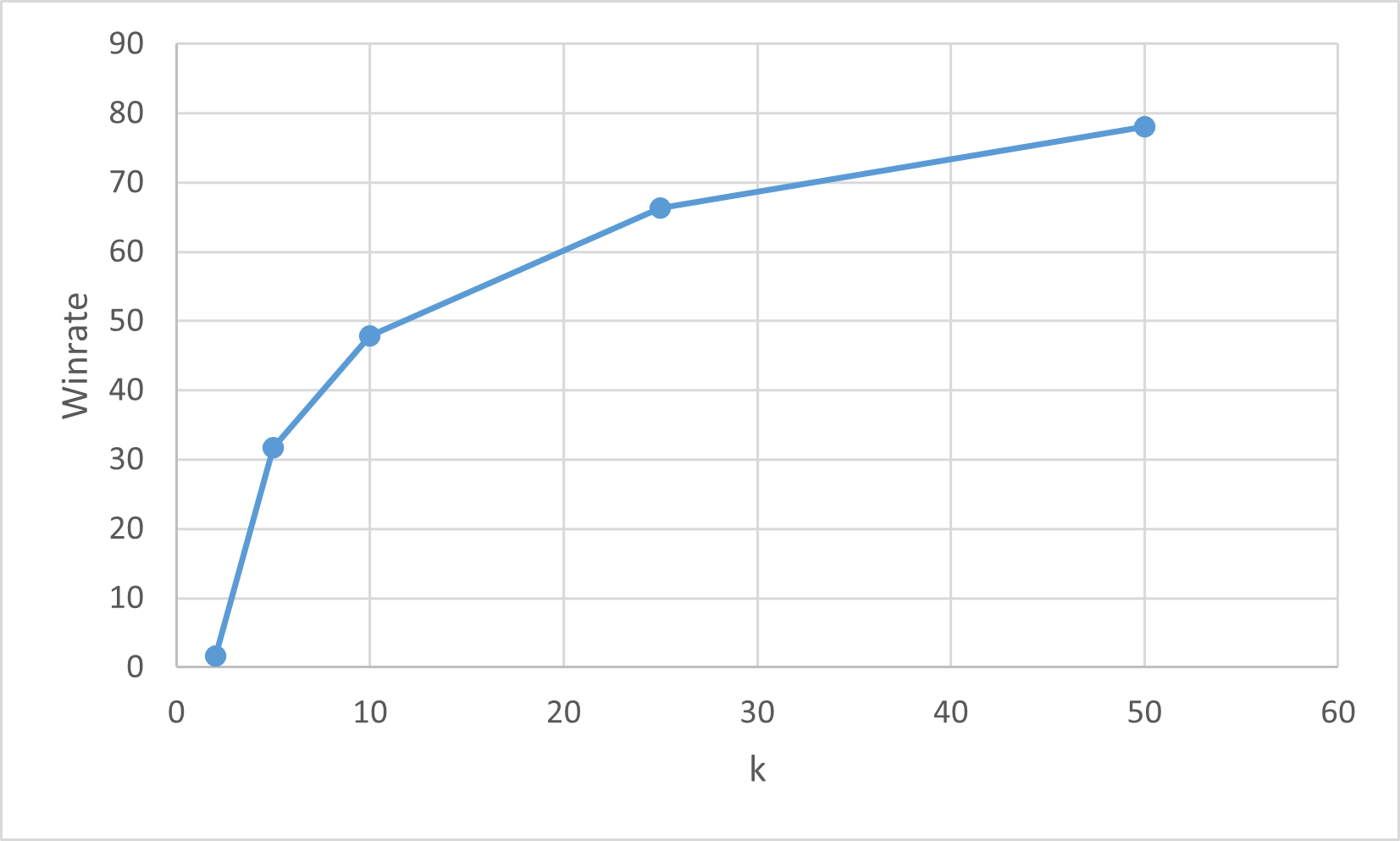}
\caption{Winrate in millichips per hand as a function of $k.$}
\label{fi:sample}
\end{figure}

Table~\ref{ta:results2} shows the results using the same parameter settings except using ``uninformed priors'' for the opponents that are independent Dirichlet with all parameters equal to 2, as done in prior work~\cite{Southey05:Bayes}. We can see that this leads to extremely poor performance, as the MBBR agent now finishes in second to last place. This shows that it is critical to use an informed prior mean strategy for the opponents, such as an exact or approximate Nash equilibrium, as opposed to a na\"ive strategy that selects actions at random. From Figure~\ref{fi:epsilon}, we can see that selection of $\epsilon$ also has a significant effect on performance, and in particular setting it too large leads to very poor performance. Figure~\ref{fi:threshold} shows that setting $H$ too large will decrease performance roughly linearly since we will miss out on rounds of exploitation. As it turns out, in this experiment values of $H$ between 0 and 200 produce very similar winrates, indicating that exploration did not play much of a role and we were able to exploit very quickly. For larger games it is likely that a larger value of $H$ is optimal. From Figure~\ref{fi:factor} we can see that changing $\eta$ leads to a negligible change in performance, with values between 1 and 8 producing winrates between 44.5 and 48.5 which decisively outperform the other agents. Finally in Figure~\ref{fi:sample} we see that the number of samples used $k$ has a very significant impact on the performance, which as expected is monotonically increasing. Larger $k$ increases the run time of the agent by a factor of $k^{n-1}$ and also increases the memory. Given the results it is clearly optimal to use as large a value of $k$ as possible given the time and memory resources available.

\begin{table}[!ht]
\centering
\begin{tabular}{|*{14}{c|}} \hline
N1 &C2 &N3 &C1 &N2 &C3 &C4 &C5 &C6 &C7 &C8 &C9 &MBBR &C10\\ \hline
47 &46 &44 &42 &40 &31 &24 &22 &19 &-19 &-25 &-42 &-54 &-175\\ \hline
\end{tabular}
\caption{Experimental results with uninformed prior.}
\label{ta:results2}
\end{table}
\normalsize
\renewcommand{\tabcolsep}{6pt}

\section{Conclusion}
\label{se:conclusion}
We presented an approach for opponent modeling in multiplayer imperfect-information games. The approach applies to the setting of partial observability of opponents' private information, which is the most general setting as it subsumes full and no observability as special cases. The approach assumes that there is no historical data or domain-specific information such as features available; therefore it is fully domain independent and applicable to any multiplayer imperfect-information game with any information revelation structure. The algorithm utilizes importance sampling and is scalable to large games. We experimented against a pool of ten real agents and three distinct exact Nash equilibrium strategies and our agent decisively came in first place. In real-world settings we often encounter unknown opponents with a wide range of strategies, and our approach leads to significantly stronger overall performance than Nash equilibrium strategies by successfully exploiting weaker opponents while also performing well against strong opponents.

\bibliographystyle{plain} 
\bibliography{C://FromBackup/Research/refs/dairefs}

\end{document}